\documentclass{article}
\usepackage{spconf,amsmath,graphicx}

\usepackage{amssymb,amsfonts}
\usepackage{graphicx}
\usepackage{textcomp}
\usepackage{xcolor}
\usepackage{cases}
\usepackage{algorithm}
\usepackage[noend]{algpseudocode}
\usepackage{tikz} 
\usepackage{cite}
\usepackage{comment}
\usepackage{balance}
\usepackage{xcolor}
\usepackage{subfigure}
\newcommand{\bs}{\boldsymbol}
\usepackage[thmmarks, amsmath, thref]{ntheorem}
 \usepackage{enumitem}

\usepackage{xspace}
\usepackage{bbm}
\usepackage{mathrsfs}

\newcommand{\Pcal}{\mathcal{P}}

 \def\E{\mathsf{E}}

\newcommand{\U}{\mathrm{Unif}}

\def\textiid{i.i.d.\@\xspace}
\newcommand\iid{\ifmmode\text{ i.i.d. } \else \textiid \fi}

\newtheorem{definition}{Definition}

\newtheorem{proposition}{Proposition}

\newcommand{\argmin}{\operatornamewithlimits{argmin}}

\usepackage[thmmarks, amsmath, thref]{ntheorem}

\title{Generalized Invariant Matching Property via Lasso}
\name{Kang Du and Yu Xiang}
\address{University of Utah\\
Electrical and Computer Engineering\\
50 Central Campus Dr \#2110, Salt Lake City, UT 84112
}
\begin{document}
\maketitle

\begin{abstract}
Learning under distribution shifts is a challenging task. One principled approach is to exploit the invariance principle via the structural causal models. However, the invariance principle is violated when the response is intervened, making it a difficult setting. In a recent work, the invariant matching property has been developed to shed light on this scenario and shows promising performance. In this work, by formulating a high-dimensional problem with intrinsic sparsity, we generalize the invariant matching property for an important setting when only the target is intervened. We propose a more robust and computation-efficient algorithm by leveraging a variant of Lasso, improving upon the existing algorithms. 
\end{abstract}
\vspace{-.5em}
\section{Introduction}
\label{sec:intro}
\vspace{-.5em}


This work concerns learning under distribution shifts, which has attracted much attention in recent years~\cite{quinonero2008dataset,weiss2016survey,csurka2017domain}. To make this challenging setting tractable, one common and principled approach is to follow the approach of using \emph{structural causal models} (SCMs)~\cite{pearl2009causality,peters2017elements} to model different data-generating mechanisms. Let $Y$ be the response given its predictors $X = (X_1,...,X_d)^\top$ in unseen environments. It is commonly assumed that the assignment $Y$ is not intervened, implying the invariant conditional distribution of $Y$ given a subset of $X$~\cite{scholkopf2012causal,zhang2013domain,peters2016causal,heinze2017conditional,buhlmann2020invariance}. Following this principle, the seminal work of invariant causal prediction (ICP)~\cite{peters2016causal} (see extensions in~\cite{heinze2018invariant,pfister2019invariant}) assumes that $\Pcal_e(Y|X_S) =  \Pcal_h(Y|X_S)$ for environments $e$ and $h$, where $(X, Y)$ is generated according to the joint distribution $\Pcal_e:=\Pcal_e^{X,Y}$, and several interesting approaches have been motivated by ICP~\cite{rojas2018invariant,magliacane2018domain,pfister2021stabilizing,rothenhausler2021anchor,arjovsky2019invariant,christiansen2021causal}.

However, it is of both theoretical and practical interest to investigate the setting when $Y$ is intervened, i.e., when the invariance principle no longer holds. This setting that allows the changes of the parameters in the assignment of $Y$ has received less attention. Recently, some progress on this setting has been made in~\cite{du2022learning} through the \emph{invariant matching property (IMP)} (preliminary results reported in~\cite{9909818}). In this work, we aim to generalize the original IMP for \emph{the important setting when only $Y$ is intervened}. First, we propose a generalized form of the IMP, which provides a unified representation of all the IMPs via a convex combination of different IMPs. This representation motivates a new perspective for identifying invariant relations from multiple environments, via learning sparse solutions in high dimensions. In particular, we show that generalized IMPs can be identified based on their intrinsic sparsity. Second, we propose an algorithm for finite samples through a variant of Lasso, which has several advantages over the existing algorithms in~\cite{du2022learning}: (1) It is more robust with respect to measurement errors and nonlinear relations (see Section~\ref{sec:exp}), (2) the algorithms from~\cite{du2022learning} identify IMPs individually, while our algorithm identifies one generalized IMP consisting of multiple IMPs directly, and (3) the shrinkage parameter of Lasso allows us to explore the trade-off between the predictive performance and generalization ability.


\vspace{-1em}
\section{Background}
\label{sec:pagestyle}
\vspace{-.5em}

Given \iid  samples $\{x_{i}^{e},y_{i}^{e}\}_{i=1}^{n_{e}}$ from each training environment $e \in \mathcal{E}^{\text{train}}$, our goal is to predict $Y$ for the \iid samples $\{x_{i}^{e^{\tau}}\}_{i=1}^{m}$ in a test environment $e^{\tau}$. We estimate a function $f:\mathbb{R}^{d} \to \mathbb{R} $ that minimizes the \emph{test population loss} $\min_{f} \E_{\Pcal_{\tau}}[l(Y,f(X))]$,
where $\Pcal_{\tau}$ denotes the distribution of $(X,Y)$ in $e^{\tau}$ and $l(y,\hat{y}) = (y-\hat{y})^2$.


A principled way to model the changes of the distribution of $(X^{e},Y^{e})$ across different environments is through linear structural causal models (SCMs). For each $e \in \mathcal{E}^{\text{all}}: = \mathcal{E}^{\text{train}}\cup \{e^{\tau}\}$, consider $(X^{e},Y^{e})$ that follows an acyclic linear SCM,
\begin{numcases}{\mathcal{M}^{e}:}
    X^{e} = \gamma Y^{e} + BX^{e}+\varepsilon_{X}^{e}  \label{eq_scm_1}\\
    Y^{e} = (\alpha^{e}+\beta)^{\top} X^{e} + \mu^{e} +\varepsilon_{Y}^{e}, \label{eq_scm_2}
\end{numcases}
where the noise distributions are invariant across environments, $\mu^{e}\in \mathbb{R}$ is a non-random shift applied on the mean of $\varepsilon_{Y}^{e}$, and the coefficient $\alpha^{e}_{j}$ is non-zero if and only if $X_{j}$ is a parent of $Y$ such that $\alpha^{e} \neq \alpha^{h}$ for some $e,h \in \mathcal{E}^{\text{all}}$ (a special case of the varying coefficient $\alpha^e$ has been studied in~\cite{du2020causal}). Note that the parameters in~\eqref{eq_scm_1} are invariant, which is a special case of the general setting in~\cite{du2022learning} that allows the parameters in both~\eqref{eq_scm_1} and~\eqref{eq_scm_2} to change. Regarding the distribution of $\varepsilon_{Y}^{e}$, we only allow its mean to change through $\mu^{e}$ and we leave the challenging setting when the variance changes for future work. The consequence of a changing variance is discussed in Corollary~2 and Remark~4 in~\cite{du2022learning}. We denote the set of parents of $Y$ with coefficients that change across environments as $PE := \{j:\alpha_{j}^{e}\neq 0\}$. The acyclic graph induced by $\mathcal{M}^{e}$ can be drawn according to the non-zero coefficients, which we denote as $\mathcal{G}(\mathcal{M}^{e})$. We briefly introduce the IMP in the next section.

\vspace{-1em}
\section{Invariant Matching Property}
\vspace{-.5em}

For $X^{e} \in \mathbb{R}^{d}$ and $Y^{e} \in \mathbb{R}$ with a joint distribution $\Pcal_{e}$, we denote the linear minimum mean squared error (LMMSE) estimator of $Y^{e}$ given $X^{e}$ by $\E_{\Pcal_{e},l}[Y|X]$. The invariant matching property (IMP) is defined as follows.

\begin{definition}[\!\cite{du2022learning}]
For $k \in \{1,\ldots,d\}$, $R \subseteq \{1,\ldots,d\}\setminus k$, and $S\subseteq \{1,\ldots,d\}$, the tuple $(k,R,S)$ is said to satisfy the invariant matching property (IMP) if, for every $e \in \mathcal{E}^{\text{all}}$, 
\begin{equation}
    \E_{\Pcal_{e},l}[Y|X_{S}]  =  \lambda \E_{\Pcal_{e},l}[X_{k}|X_{R}] + \eta^{\top}X^{e} \label{imp}
\end{equation}
holds for some $\lambda$ and $\eta$ that are invariant. 
\end{definition}
\vspace{-.5em}

Under different intervention settings, several classes of IMPs are characterized in~\cite{du2022learning}. Interestingly, the characterized IMPs will imply an alternative form of invariance, when $Y$ is intervened, represented as follows: $\Pcal_e(Y|\phi_{e}(X)) =  \Pcal_h(Y|\phi_{h}(X))$ holds for all $e,h \in \mathcal{E}^{\text{all}}$ (see~\cite{du2022learning} for details). This idea of invariance will not be pursued in this work, since we focus on using the IMP directly for the prediction of $Y$. Based on a natural decomposition of the IMP (see Section~3.3 in~\cite{du2022learning}), Theorem~1 from~\cite{du2022learning} provides sufficient conditions for IMPs to hold, which implies the following proposition.

\begin{proposition}[\!\cite{du2022learning}]\label{prop:exists}
Assume that $Y$ has at least one child in $\mathcal{G}({\mathcal{M}^{e}})$, the IMP holds for any $k \not\in PE$ and $R \supseteq PE$, and $S =\{1,\ldots,d\}$, if the coefficients of $\E_{\Pcal_{e},l}[X_{k}|X_{R}]$ are not invariant across environments.
\end{proposition}
\vspace{-.5em}

For the setting considered in $\mathcal{M}^{e}$, since an IMP with $S=\{1,\ldots,d\}$ is optimal for the prediction of $Y^{e}$ among all linear functions of $X^{e}$, we refer to IMP as IMP with $X_{S}=X$. Note that IMPs may not hold for $X_{S}=X$ in general when $\gamma$ and $B$ also depend on $e$ or the distribution of $e_{X}$ changes with $e$ (see Section~4.2 from~\cite{du2022learning}). For any tuple $(k,R)$, the feature $\E_{\Pcal_{e},l}[X_{k}|X_{R}]$ is also called a prediction module in~\cite{du2022learning}. According to whether the IMP is satisfied and whether a prediction module is an invariant linear function of $X^{e}$, we further classify the prediction modules into three categories.


\begin{definition}
\begin{enumerate}[leftmargin=*,itemsep=01pt,parsep=1pt]
\item \emph{Matched prediction module}: a module that satisfies an IMP.
    \item \emph{Redundant prediction module}:  a module that does not satisfy any IMPs and its coefficients are invariant across environments. 
    \item \emph{Anti-matching prediction module}: a module that does not satisfy any IMPs and its coefficients change with environments. 
\end{enumerate}
\end{definition}
\vspace{-.5em}

\begin{figure}[h]
\centering
\vspace{-1em}
\includegraphics[width=0.6\linewidth]{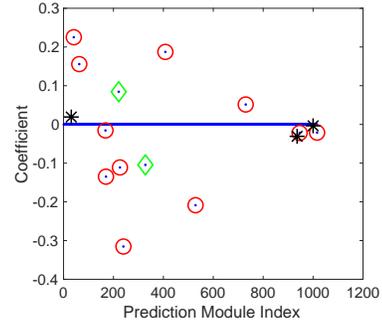}
\caption{One selection result by our algorithm, where each blue dot is one prediction module. The matched modules,  redundant modules, and anti-matching modules are marked by the red circle, green diamond, and black star, respectively.}
\label{ill_selection}
\vspace{-2em}
\end{figure}

The role of matched prediction modules is to capture the changes of the parameters $\alpha^{e}$ and $\mu^{e}$, so that $\E_{\Pcal_{e},l}[Y|X]$ can be represented as an invariant linear function. Since a redundant prediction module is an invariant linear function of $X^{e}$, it is redundant given $X^{e}$. Apparently, when the distribution of $(X_{k}^{e},X_{R}^{e})$ is invariant across environments, the corresponding prediction module is redundant. To be specific, for any $k \not \in CH(Y)$ and $R=PA(X_{k})$, the prediction module $\E_{\Pcal_{e},l}[X_{k}|X_{R}]= B_{k,\cdot}X^{e}$ is redundant. When the distribution of $(X_{k}^{e},X_{R}^{e})$ is not invariant, the redundancy  happens when $X_{k}$ and $X_{R}$ are independent and $\E_{\Pcal_{e}}[X_{k}]$ is invariant, in which case $\E_{\Pcal_{e},l}[X_{k}|X_{R}]= \E_{\Pcal_{e}}[X_{k}]$. For prediction modules that are not redundant, examples of anti-matching prediction modules can be found by reversing the conditions in Proposition~\ref{prop:exists}, i.e., $k \in PE$ and $R \not \supseteq PE$. Also, see the motivating example in~\cite{du2022learning} with two matched prediction modules and one anti-matching prediction module with $ k \in PE$. In the next two sections, we propose a generalized form of the IMP and an algorithm for identifying the generalized IMP. The algorithm shows a trend of selecting matched prediction modules while avoiding anti-matching prediction modules (as illustrated in Fig.~\ref{ill_selection}), where the matched modules have larger (in absolute value) coefficients in comparison to the anti-matching modules.

\vspace{-.5em}
\section{A Generalization of the IMP} \label{sec:general_imp}
\vspace{-.5em}

For the finite samples implementation, the main task is to identify matched prediction modules. Recall that when $S$ in the IMP is fixed to be $\{1,\ldots,d\}$, then a straightforward way is to search exhaustively over all $(k,R)$'s and test the null hypothesis that the IMP holds for each $(k,R)$ as in~\cite{du2022learning}. However, this procedure can be sensitive with respect to violations of the model assumptions, e.g., nonlinear relations and measurement errors, since there could be no IMPs in these settings. To this end, we propose a generalized form of the IMP method that is equivalent to a high-dimensional linear model with intrinsic sparsity. It turns out we can make use of sparsity to identify the generalized invariant relation,  leading to a more robust procedure detailed in the next section.  


First, since IMPs are not unique in general, a convex combination of $q\geq 2$ different IMPs will result in the following invariant relation, 
\begin{align}
        \E_{\Pcal_{e},l}[Y|X]  &= \sum_{i=1}^{q} a_{i}\left( \lambda^{(i)} \E_{\Pcal_{e},l}[X_{k_{i}}|X_{R_{i}}] + \left(\eta^{(i)}\right)^{\top}X^{e}\right),  \nonumber
\end{align}  
with $0 < a_{1},\ldots,a_{q} < 1$ and $\textstyle\sum_{i=1}^{q}a_{i}=1$. This motivates a generalized form of the IMP method.
\begin{definition}
The variables $X^{e}$ and $Y^{e}$ are said to satisfy the \emph{generalized invariant matching property} if, for every $e \in \mathcal{E}^{\text{all}}$, 
\begin{equation}
      \E_{\Pcal_{e},l}[Y|X] = \theta^{\top}Z^{e} + \zeta^{\top}X^{e}, \label{eq:convx_comb_inp}
\end{equation}
holds for some $\theta \in \mathbb{R}^{p}$, $\zeta \in \mathbb{R}^{d}$ that are invariant across environments, and $Z^{e}=[Z_1^e,...,Z_p^e]^\top \in \mathbb{R}^{p}$ is the vector of all the prediction modules with dimension $p= d (2^{d-1}-1)$.
\end{definition}

 Observe that~\eqref{eq:convx_comb_inp} reduces to the IMP when only one of the prediction modules has a nonzero coefficient. We provide some insights into the generalized IMP from a different perspective. Define the population linear regression model,
\begin{equation}
      Y^{e} =\theta^{\top}Z^{e} + \zeta^{\top}X^{e} + \varepsilon^{e}. \label{eq:high_lm}
\end{equation}
Observe that~\eqref{eq:convx_comb_inp} is equivalent to~\eqref{eq:high_lm} with $\varepsilon^{e} = Y^{e} - \E_{\Pcal_{e},l}[Y|X] $. In other words,~\eqref{eq:convx_comb_inp} is satisfied if and only if 
\begin{equation}
  (\theta,\zeta) \in \argmin_{(a,b)}\;\E_{\Pcal_{e}}[(Y-a^{\top}Z- b^{\top}X)^{2}], \label{eq:mse_hlm}
\end{equation}
for every $e \in \mathcal{E}^{\text{all}}$. This is due to the definition of LMMSE and the fact that $a^{\top}Z- b^{\top}X$ is a linear function of $X$ (recall that each prediction module $Z_{j}$ is a linear function of $X$).

In the finite-sample setting, as the size of the graph $d$ grows, the objective function in~\eqref{eq:mse_hlm} will quickly become a high-dimensional problem, namely, the sample size $n$ becomes much smaller compared to the dimension $p+d$, making the problem ill-posed. Fortunately, if there exists at least one IMP, any solution to~\eqref{eq:mse_hlm} under the constraint $||a||_{0}:=\#\{j\geq 1:a_{j}\neq 0\} =  1$ will lead to an IMP, following from the fact that the generalized IMP reduces to the IMP under the sparsity constraint $||\theta||_{0}=1$. Moreover, the sparsity constraint will make~\eqref{eq:mse_hlm} well-defined in finite samples. Following this idea, it is natural to impose a sparsity constraint $||a|||_{0}:=\#\{j\geq 1:a_{j}\neq 0\} =  s$ such that $s \ll n$ to help identify generalized IMPs. In practice, instead of solving~\eqref{eq:mse_hlm} with a $l_{0}$ norm constraint, we propose a relaxed version using a variant of Lasso in the next section.

\vspace{-1em}
\section{Learning Generalized IMP via Lasso}
\label{sec:learn_gimp}
\vspace{-.5em}

Consider a set of training environments $\mathcal{E}^{\text{train}} = \{e_{1},\ldots,e_{s}\}$ and data matrices $\bs{X}_{e_{i}} \in \mathbb{R}^{n_{e_{i}} \times d}$ and  $\bs{Y}_{e_{i}} \in \mathbb{R}^{n_{e_{i}}}$ for $e_{i} \in \mathcal{E}^{\text{train}} $. The pooled data matrices are denoted by $\bs{X} \in \mathbb{R}^{n \times d}$ and $\bs{Y} \in \mathbb{R}^{n}$, where $n =\sum_{i=1}^{s}n_{e_{i}}$. We assume that the problem is in low-dimensional (i.e., $d \ll n$) or the predictors have been pre-selected in high dimensions. Then, we defined a data matrix with columns being all the estimated prediction modules as $\bs{\hat{Z}} \in \mathbb{R}^{n \times p}$, where each of its columns is computed using OLS in different environments as follows,
\begin{equation}\label{eq:esti_xp}
    \bs{\hat{Z}}_{j} = \begin{bmatrix}
    \bs{X}_{e_{1},R}(\bs{X}_{e_{1},R}^{\top}\bs{X}_{e_{1},R})^{-1}\bs{X}_{e_{1},R}^{\top}\bs{X}_{e_{1},k}\\
     \vdots\\
      \bs{X}_{e_{s},R}(\bs{X}_{e_{s},R}^{\top}\bs{X}_{e_{s},R})^{-1}\bs{X}_{e_{s},R}^{\top}\bs{X}_{e_{s},k}
    \end{bmatrix}
\end{equation}
with each $j\in\{1,\ldots,p\}$ corresponding to a unique tuple $(k,R)$. The corresponding matrix of the true prediction modules is denoted by $\bs{Z} \in \mathbb{R}^{n\times p}$ (obtained by replacing the OLS estimates with the population OLS parameters). Under the assumptions in Proposition~\ref{prop:exists}, there exists a generalized IMP with some parameters $\{\theta,\zeta\}$. Then, according to~\eqref{eq:high_lm},
\begin{equation}
    \bs{Y}  = \bs{Z}\theta + \bs{X}\zeta + \bs{\varepsilon},
 \label{eq:imp_vector}
 \end{equation}
 where~$\bs{\varepsilon}: = \bs{Y} - \bs{\hat{Y}}$. The vector $\bs{\hat{Y}} \in \mathbb{R}^{n}$ is defined by stacking up the vectors $\boldsymbol{X}_{e_{i}}\beta^{(e_i)}$'s, $e_i \in\mathcal{E}^{\text{train}}$, where $\beta^{(e_i)}$ is the population OLS estimator when regressing $Y^{e_{i}}$ on $X^{e_{i}}$. 


\begin{figure*}[t]\centering
\begin{subfigure}{}\includegraphics[scale=0.19]{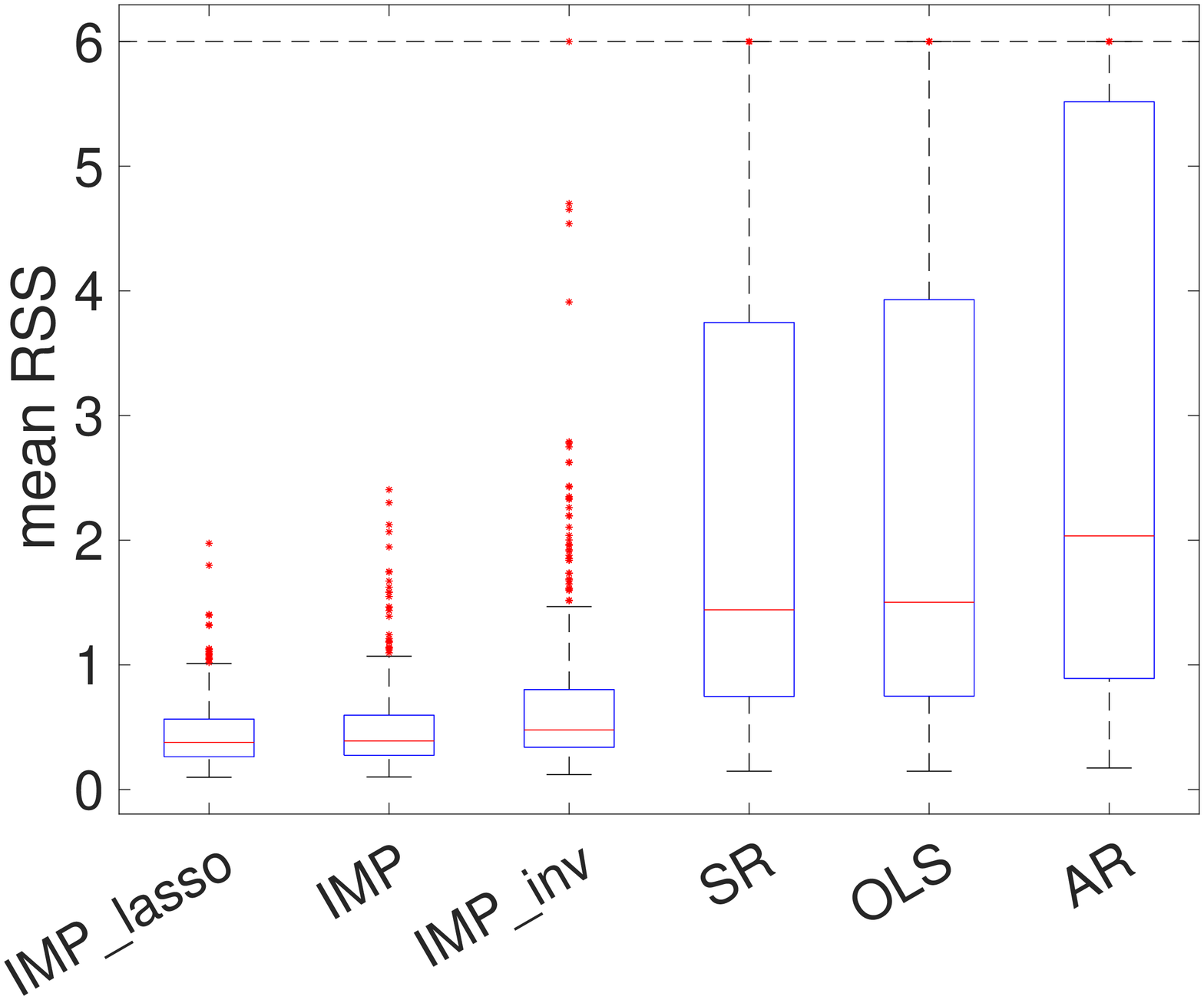}\label{exp6_1}
\end{subfigure}
\begin{subfigure}{}\includegraphics[scale=0.19]{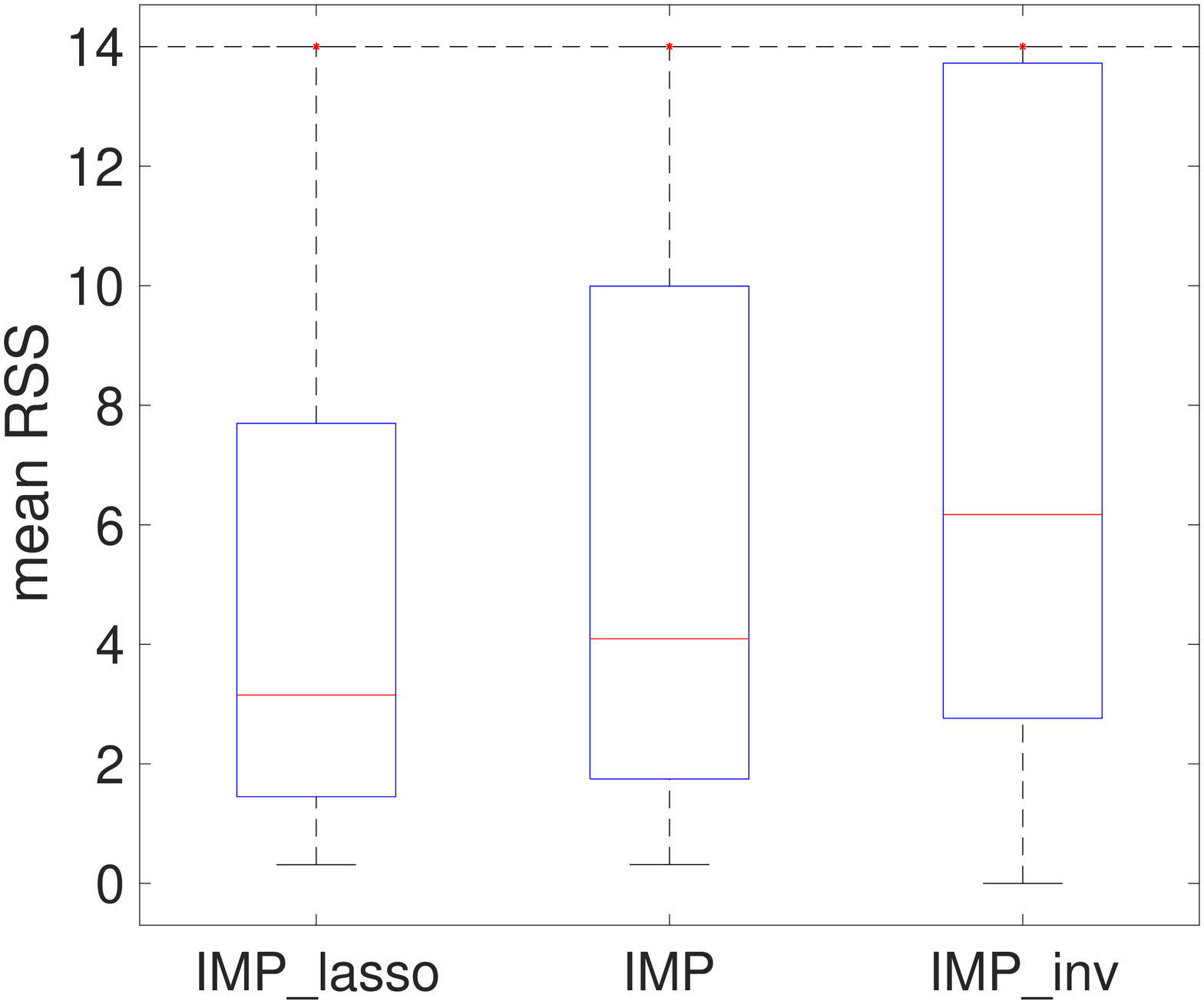}
\end{subfigure}
\begin{subfigure}{}\includegraphics[scale=0.19]{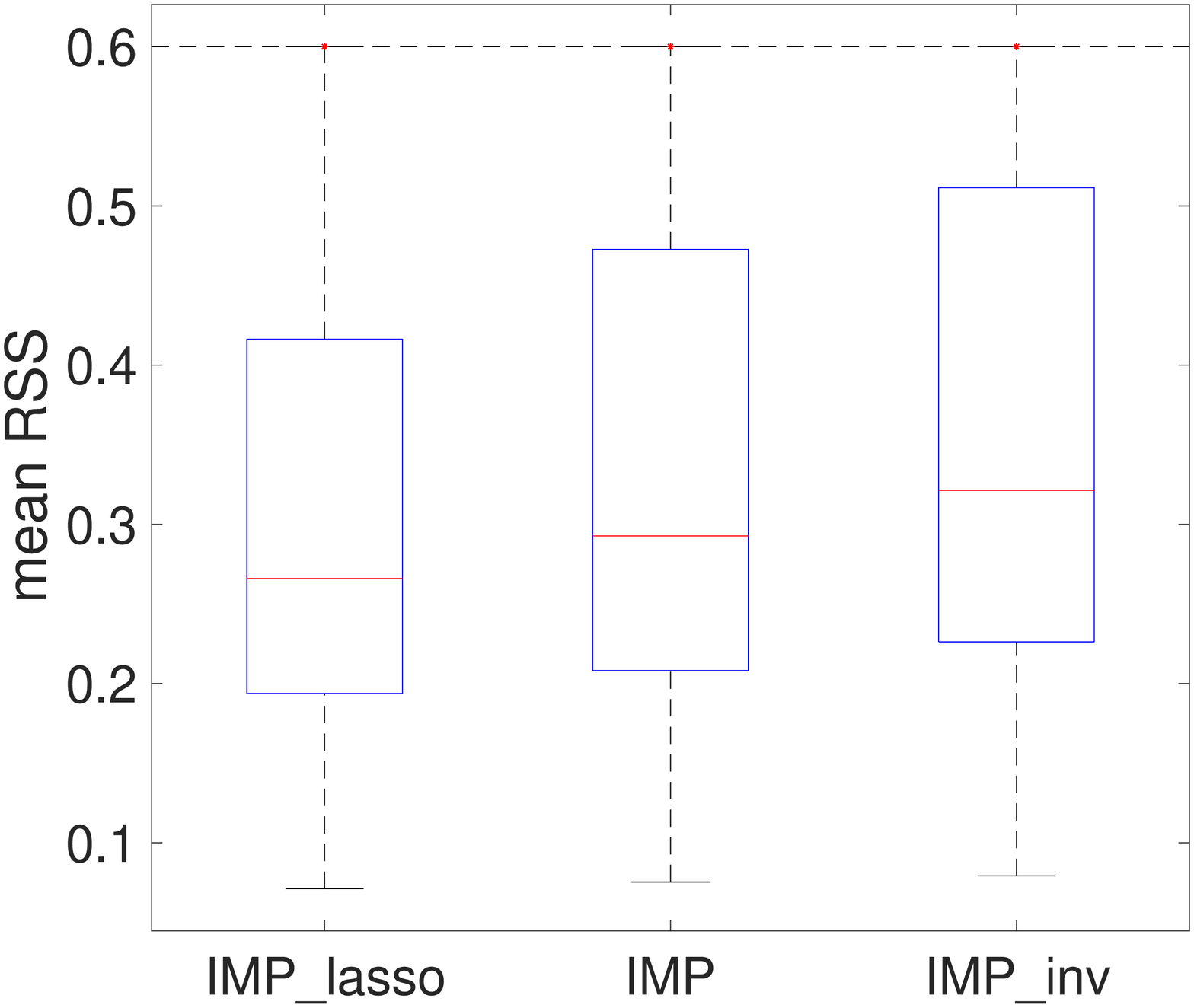}
\end{subfigure}
\vspace{-1em}
\caption{ From left to right, the three figures (a), (b), and (c) correspond to Expt. 6.1, Expt. 6.2.1, and Expt. 6.2.2, respectively.}
\vspace{-1em}
\end{figure*}

Motivated by the observations from Section~\ref{sec:general_imp}, we propose to learn a generalized IMP through the following penalized regression that promotes sparsity,
\begin{equation}
  \min_{\theta \in \mathbb{R}^{p},\zeta \in \mathbb{R}^{d}}  ||\boldsymbol{Y} - \boldsymbol{\hat{Z}}\theta-\boldsymbol{X}\zeta||^{2} + \lambda ||\theta||_{1} \label{eq:penal_regress}
\end{equation}
for $\lambda >0$. 
Except for $\boldsymbol{\hat{Z}}$ being an estimate,~\eqref{eq:penal_regress} is a variant of the Lasso problem with a sparsity constraint on only a subset of the coefficients ($\theta$ instead of $(\theta, \zeta)$). The following proposition draws a connection between~\eqref{eq:penal_regress} and the Lasso~\cite{tibshirani1996regression}. 

\begin{proposition}\label{prop:equi_lasso}
Assume that~$\bs{X}$ has full rank, the solution to~\eqref{eq:penal_regress} denoted by $(\theta^{*},\zeta^{*})$ can be represented as
\begin{equation}
 \theta^{*} \in  \argmin_{\theta \in \mathbb{R}^{p}}  ||\bs{Y}' - \bs{X}'\theta||^{2} + \lambda ||\theta||_{1} \label{eq:equiv_lasso}
\end{equation}
with $\bs{Y}' =(I-P)\bs{Y}$ and $\bs{X}' = (I-P)\bs{\hat{Z}}$ for the projection matrix $P =\bs{X} (\bs{X}^{\top}\bs{X})^{-1}\bs{X}^{\top}$, and
\begin{equation}
   \zeta^{*} = (\bs{X}^{\top}\bs{X})^{-1}\bs{X}^{\top}(\boldsymbol{Y} - \boldsymbol{\hat{Z}}\theta^{*}). 
\end{equation}
\end{proposition}

\emph{\noindent Proof: Let $l(\theta,\zeta)$ denote the objective function of~\eqref{eq:penal_regress}. Assume that $\bs{X}$ has full rank, then $l(\theta,\zeta)$ is strictly convex with respect to $\zeta$. But we do not assume that $\bs{\hat{Z}}$ has full rank, thus $l(\theta,\zeta)$ is convex with respect to $\theta$ but may not be strictly convex in general. Based on these facts, the solution to~\eqref{eq:penal_regress} can be found by minimizing over $\zeta$ first and then $\theta$. First, for any fixed $\theta$, minimizing $l(\theta,\zeta)$ over $\zeta$ leads to
\begin{equation}
    \zeta^{*}(\theta) = (\bs{X}^{\top}\bs{X})^{-1}\bs{X}^{\top}(\boldsymbol{Y} - \boldsymbol{\hat{Z}}\theta) \nonumber
\end{equation}
 that depends on $\theta$. Then, minimizing $l(\theta,\zeta^{*}(\theta))$ over $\theta$ is simply~\eqref{eq:equiv_lasso}. Finally, $\zeta^{*}$ is given by $\zeta^{*}(\theta^{*})$. }

\smallskip


There are several benefits of solving the relaxed problem~\eqref{eq:penal_regress} instead of identifying IMPs one by one as in~\cite{du2022learning}. First, since IMPs are not unique in general, it is not clear how to make use of all the identified IMPs. In~\cite{du2022learning}, the authors choose to take an average (with equal weights) over all the identified IMPs, however, assigning equal weights to different IMPs can be restrictive, whereas~\eqref{eq:penal_regress} provides a more flexible way of making use of multiple IMPs. Note that the coefficients $\theta_{j}$'s will determine which prediction modules are more important for the prediction of $Y$. Second, the linear SCM assumption may be violated in practical settings (e.g., under measurement errors or nonlinear relations), so that IMPs can only hold approximately and~\eqref{eq:penal_regress} allows us to make use of the approximated IMPs for the prediction of $Y$. Third, the shrinkage parameter $\lambda$ allows us to explore the trade-off between the prediction performance and generalization ability. For the general setting when all the parameters in $\mathcal{M}^{e}$ are allowed to change across environments (i.e., when $X_{S} \neq X$ in the IMP), there is a potential benefit of our formulation in terms of computation efficiency, since the algorithms from~\cite{du2022learning} have to deal with an exhaustive search over all $(k,R,S)'s$. When both $X$ and $Y$ are intervened, however, more sophisticated regularizers beyond sparsity are needed.





\vspace{-1em}
\section{Experiment}
\label{sec:exp}
\vspace{-1em}

We compare our method named $\text{IMP}_{\text{lasso}}$  with $5$ baselines below: IMP~\cite{du2022learning}, $\text{IMP}_{\text{inv}}$~\cite{du2022learning}, OLS, stabilized regression (SR)~\cite{pfister2021stabilizing}, and anchor regression (AR)~\cite{rothenhausler2021anchor}. There are other methods that are compared in~\cite{du2022learning}, while their performances are less competitive. Thus, we do not compare our method with those ones. For IMP and $\text{IMP}_{\text{inv}}$, we fix $S= \{1,\ldots,d\}$ since we focus on the setting when only the parameters in the assignment of $Y$ change. The significance levels of all the methods are fixed to be $0.05$. The shrinkage parameter $\lambda$ of Lasso is selected using a $5$-fold cross validation. The predictive performance is measured by the mean residual sum of squares (RSS) on the test data.

\noindent\underline{\bf 6.1 Regular settings.} We follow the same experiment setting as Section 7.1.2 from~\cite{du2022learning}. For each data set, we first randomly generate a linear SCM without varying parameters, where the acyclic graph has $9$ node and the graph structure is determined by a low-triangular matrix of \iid Bernoulli(1/2) random variables. Then, we randomly select one node as $Y$ while requiring that $Y$ has at least one child and one parent. The nonzero coefficients in the linear SCM are sampled from $\U[-0.5,-1.5]\cup[0.5,1.5]$, and the noise variables are standard normal. To introduce the parameters that change across environments in the assignment of $Y$, we add perturbation terms to the original parameters in each environment. Consider training environments $\mathcal{E}^{\text{train}}=\{1,\ldots,5\}$ and test environments  $\mathcal{E}^{\text{test}}=\{6,\ldots,10\}$. We choose the set $PE$ by randomly selecting $n_{p} \sim \U\{1,\ldots,|PA(Y)|\}$ of parents of $Y$ to have varying coefficients. For each $j \in PE$, we add a perturbation term sampled from $\U[-a,a]$ to the original coefficient of $X_{j}$ in the assignment of $Y$ in each environment, where $a=2$ for the training environment and $a=10$ for the test environments. The shift on the mean of the noise of $Y$ is added in the same way as the perturbation on the coefficients. For each experiment, the sample size is $300$ for each environment and we simulate $500$ data sets. From Fig.~2(a), our method performs similarly to the original IMP from~\cite{du2022learning} by having smaller median and variance of the mean RSS. 


\noindent\underline{\bf Robustness.} Since other baselines are not as competitive as the IMP-based methods when the parameters in the assignment of $Y$ change, we only compare with $\text{IMP}$ and $\text{IMP}_{\text{inv}}$ for the experiments on robustness.

    
\noindent\underline{\bf 6.2.1 Robustness: measurement error.} In the linear SCM, we add \iid error terms that are $\mathcal{N}(0,\sigma^{2})$-distributed to all the variables. But note that we do not add the error term to $Y$ in the test environments, since $Y$ is not observed. The results for $\sigma^{2}=2.5$ (see Fig.~2(b)) show that our method is more robust against measurement errors compared with $\text{IMP}$ and $\text{IMP}_{\text{inv}}$. As~$\sigma^{2}$ increases, the gap between our method and the baselines can be even larger, but the performances of all three methods are getting worse.


\noindent\underline{\bf 6.2.2 Robustness: nonlinear relations.} We introduce nonlinearities by adding a subsequent nonlinear transform to the assignment of $Y$. The transform is defined by $f(x)= \text{sign}(x) |x|^{b} $. For $b=0.5$, the results from Fig.~2(c) show the improved robustness of our method compared with $\text{IMP}$ and $\text{IMP}_{\text{inv}}$. As $b$ gets deviated from $1$, the robustness of our method can disappear, since linear relations can be poor approximates of highly nonlinear functions.

\newpage
\balance
\bibliographystyle{IEEEtran}
\bibliography{ref.bib}

\begin{thebibliography}{10}
\providecommand{\url}[1]{#1}
\csname url@samestyle\endcsname
\providecommand{\newblock}{\relax}
\providecommand{\bibinfo}[2]{#2}
\providecommand{\BIBentrySTDinterwordspacing}{\spaceskip=0pt\relax}
\providecommand{\BIBentryALTinterwordstretchfactor}{4}
\providecommand{\BIBentryALTinterwordspacing}{\spaceskip=\fontdimen2\font plus
\BIBentryALTinterwordstretchfactor\fontdimen3\font minus
  \fontdimen4\font\relax}
\providecommand{\BIBforeignlanguage}[2]{{%
\expandafter\ifx\csname l@#1\endcsname\relax
\typeout{** WARNING: IEEEtran.bst: No hyphenation pattern has been}%
\typeout{** loaded for the language `#1'. Using the pattern for}%
\typeout{** the default language instead.}%
\else
\language=\csname l@#1\endcsname
\fi
#2}}
\providecommand{\BIBdecl}{\relax}
\BIBdecl

\bibitem{quinonero2008dataset}
J.~Quinonero-Candela, M.~Sugiyama, A.~Schwaighofer, and N.~D. Lawrence,
  \emph{Dataset shift in machine learning}.\hskip 1em plus 0.5em minus
  0.4em\relax Mit Press, 2008.

\bibitem{weiss2016survey}
K.~Weiss, T.~M. Khoshgoftaar, and D.~Wang, ``A survey of transfer learning,''
  \emph{Journal of Big Data}, vol.~3, no.~1, pp. 1--40, 2016.

\bibitem{csurka2017domain}
G.~Csurka, ``Domain adaptation for visual applications: A comprehensive
  survey,'' \emph{arXiv preprint arXiv:1702.05374}, 2017.

\bibitem{pearl2009causality}
J.~Pearl, \emph{Causality}.\hskip 1em plus 0.5em minus 0.4em\relax Cambridge
  University Press, 2009.

\bibitem{peters2017elements}
J.~Peters, D.~Janzing, and B.~Sch{\"o}lkopf, \emph{Elements of causal
  inference: foundations and learning algorithms}.\hskip 1em plus 0.5em minus
  0.4em\relax The MIT Press, 2017.

\bibitem{scholkopf2012causal}
B.~Sch{\"o}lkopf, D.~Janzing, J.~Peters, E.~Sgouritsa, K.~Zhang, and J.~Mooij,
  ``On causal and anticausal learning,'' in \emph{29th International Conference
  on Machine Learning (ICML 2012)}.\hskip 1em plus 0.5em minus 0.4em\relax
  International Machine Learning Society, 2012, pp. 1255--1262.

\bibitem{zhang2013domain}
K.~Zhang, B.~Sch{\"o}lkopf, K.~Muandet, and Z.~Wang, ``Domain adaptation under
  target and conditional shift,'' in \emph{International Conference on Machine
  Learning}.\hskip 1em plus 0.5em minus 0.4em\relax PMLR, 2013, pp. 819--827.

\bibitem{peters2016causal}
J.~Peters, P.~B{\"u}hlmann, and N.~Meinshausen, ``Causal inference by using
  invariant prediction: identification and confidence intervals,''
  \emph{Journal of the Royal Statistical Society. Series B (Statistical
  Methodology)}, pp. 947--1012, 2016.

\bibitem{heinze2017conditional}
C.~Heinze-Deml and N.~Meinshausen, ``Conditional variance penalties and domain
  shift robustness,'' \emph{arXiv preprint arXiv:1710.11469}, 2017.

\bibitem{buhlmann2020invariance}
P.~B{\"u}hlmann, ``Invariance, causality and robustness,'' \emph{Statistical
  Science}, vol.~35, no.~3, pp. 404--426, 2020.

\bibitem{heinze2018invariant}
C.~Heinze-Deml, J.~Peters, and N.~Meinshausen, ``Invariant causal prediction
  for nonlinear models,'' \emph{Journal of Causal Inference}, vol.~6, no.~2,
  2018.

\bibitem{pfister2019invariant}
N.~Pfister, P.~B{\"u}hlmann, and J.~Peters, ``Invariant causal prediction for
  sequential data,'' \emph{Journal of the American Statistical Association},
  vol. 114, no. 527, pp. 1264--1276, 2019.

\bibitem{rojas2018invariant}
M.~Rojas-Carulla, B.~Sch{\"o}lkopf, R.~Turner, and J.~Peters, ``Invariant
  models for causal transfer learning,'' \emph{The Journal of Machine Learning
  Research}, vol.~19, no.~1, pp. 1309--1342, 2018.

\bibitem{magliacane2018domain}
S.~Magliacane, T.~Van~Ommen, T.~Claassen, S.~Bongers, P.~Versteeg, and J.~M.
  Mooij, ``Domain adaptation by using causal inference to predict invariant
  conditional distributions,'' \emph{Advances in Neural Information Processing
  Systems}, vol.~31, 2018.

\bibitem{pfister2021stabilizing}
N.~Pfister, E.~G. Williams, J.~Peters, R.~Aebersold, and P.~B{\"u}hlmann,
  ``Stabilizing variable selection and regression,'' \emph{The Annals of
  Applied Statistics}, vol.~15, no.~3, pp. 1220--1246, 2021.

\bibitem{rothenhausler2021anchor}
D.~Rothenh{\"a}usler, N.~Meinshausen, P.~B{\"u}hlmann, and J.~Peters, ``Anchor
  regression: Heterogeneous data meet causality,'' \emph{Journal of the Royal
  Statistical Society: Series B (Statistical Methodology)}, vol.~83, no.~2, pp.
  215--246, 2021.

\bibitem{arjovsky2019invariant}
M.~Arjovsky, L.~Bottou, I.~Gulrajani, and D.~Lopez-Paz, ``Invariant risk
  minimization,'' \emph{arXiv preprint arXiv:1907.02893}, 2019.

\bibitem{christiansen2021causal}
R.~Christiansen, N.~Pfister, M.~E. Jakobsen, N.~Gnecco, and J.~Peters, ``A
  causal framework for distribution generalization,'' \emph{IEEE Transactions
  on Pattern Analysis and Machine Intelligence}, 2021.

\bibitem{du2022learning}
K.~Du and Y.~Xiang, ``Learning invariant representations under general
  interventions on the response,'' \emph{arXiv preprint arXiv:2208.10027},
  2022.

\bibitem{9909818}
------, ``An invariant matching property for distribution generalization under
  intervened response,'' in \emph{2022 30th European Signal Processing
  Conference (EUSIPCO)}, 2022, pp. 1387--1391.

\bibitem{du2020causal}
------, ``Causal inference from slowly varying nonstationary processes,''
  \emph{arXiv preprint arXiv:2012.13025}, 2020.

\bibitem{tibshirani1996regression}
R.~Tibshirani, ``Regression shrinkage and selection via the lasso,''
  \emph{Journal of the Royal Statistical Society: Series B (Methodological)},
  vol.~58, no.~1, pp. 267--288, 1996.

\end{thebibliography}
\end{document}